\documentclass[onecolumn,12pt,prd]{revtex4}
\usepackage{epsfig,color,rotating,lineno,url}

\graphicspath{{Figures/}}

\begin{document}

\title{
  \large \bf \boldmath
  A Method to Measure $\cos(2\beta)$ 
  Using Time-Dependent Dalitz Plot Analysis of $B^0 \to D_{CP} \pi^+\pi^-$
}

\author{Thomas Latham}
\author{Tim Gershon}
\affiliation{Department of Physics, University of Warwick, Coventry CV4 7AL, United Kingdom}

\date{\today}

\begin{abstract}
We present a feasibility study of a previously outlined method to measure both
the sine and the cosine of twice the CKM Unitarity Triangle angle $\beta$
using a time-dependent Dalitz plot analysis of $B^0 \to D \pi^+\pi^-$ decays,
where the neutral $D$ meson is reconstructed in decays to $CP$ eigenstates.  
We show that this method can
be used at the $B$ factories to make a measurement of $\cos(2\beta)$ that is
competitive with, or more precise than, other techniques using different
quark-level transitions, while $\sin(2\beta)$ can be measured to a better
precision than any existing measurement using $b \to c \bar{u} d$
transitions. Furthermore, this technique has great potential to be employed at
LHCb.
\end{abstract}

\maketitle

\setcounter{footnote}{0}

\newpage

\section{Introduction}
\label{sec:intro}

Precise determinations of the elements of the Cabibbo-Kobayashi-Maskawa
(CKM)~\cite{Cabibbo:1963yz,Kobayashi:1973fv}
matrix are important to check the consistency of the Standard Model and to
search for new physics.
Measurements of the $B^0 - \bar{B}^0$ mixing phase determined from
$B^0 \to J/\psi K^0$ (and similar) decays at the $B$ factories give
$\sin(2\beta) = 0.680 \pm
0.025$~\cite{Aubert:2007hm,Chen:2006nk,Barberio:2008fa}, where $\beta$ is one
of the angles of the CKM Unitarity Triangle (for an introduction, see, for
example~\cite{Harrison:1998yr,Bigi:2000yz,Branco:1999fs}).
These results confirm the Kobayashi-Maskawa mechanism as the origin of $CP$
violation within the Standard Model.  Nonetheless, the effects of ``new
physics'' are expected to be seen as non-negligible corrections to the
Standard Model, particularly if new particles are present at energies as low
as the TeV scale.  The purpose of flavour physics in the LHC era, as discussed
in several recent
reviews~\cite{delAguila:2008iz,Buchalla:2008jp,Raidal:2008jk,Bona:2007qt,Gershon:2006mt,Browder:2007gg,Browder:2008em},
is to constrain the new physics parameter space and -- once observed -- to
measure the couplings of the new physics particles.

An important part of this programme is the precise measurement of the angles
of the CKM Unitarity Triangle using different quark-level transitions.  For
example, comparisons of the value of $\sin(2\beta)$ measured in $b \to
c\bar{c}s$ transitions (such as $B^0 \to J/\psi K^0$) with those obtained in 
penguin (loop) dominated $b \to s\bar{s}s$ transitions (such as $B^0 \to \phi
K^0$) or $b \to c\bar{u}d$ transitions ({\it e.g.} $B^0 \to D\pi^0$) can probe
for new physics
effects~\cite{Grossman:1996ke,Fleischer:1996bv,London:1997zk,Ciuchini:1997zp}.
In order to achieve the best possible precision, as well as to remove
ambiguities in the results, it is important to use channels that can measure
$\cos(2\beta)$ as well as $\sin(2\beta)$.  This can be achieved, in general,
by using any final state that contains interfering amplitudes.

In this paper, we present the results of a study of the feasibility of
measuring $\cos(2\beta)$ from $B^0 \to D \pi^+\pi^-$ decays, where the neutral
$D$ meson is reconstructed in decays to $CP$ eigenstates.  This method has
previously been described in outline~\cite{Charles:1998vf}.  In this work, we
significantly extend the earlier study, taking advantage of results from the
$B$ factories that provide information on the composition of the Dalitz
plot~\cite{Abe:2003zm,Kuzmin:2006mw}. 
We estimate the sensitivity that can be achieved with the BaBar dataset, and
comment on the potential of the LHCb experiment.

The paper is organised as follows.
In Section~\ref{sec:review} we briefly review alternative approaches to
measure $\cos(2\beta)$.
The main body of the paper is in Sections~\ref{sec:method}
and~\ref{sec:feasibility}, in which we give a description of the method and
discuss the results of our feasibility study.
Finally, we present our conclusions.

\clearpage
\section{Review}
\label{sec:review}

The weak phase $2\beta$ can be probed through mixing-induced $CP$ violation
effects in $B$ decays mediated by a number of different quark-level
transitions.  This enables a powerful test of the Standard Model, since models
of physics beyond the Standard Model that introduce new particles at the TeV
scale can produce effects that differ between different decay modes.
In this section we briefly review the various techniques that
have been suggested, in order to illustrate the need for methods that can be
used to provide precise measurements at LHCb and future experiments.

\underline{$b \to c \bar{c} s$}

Since the most experimentally precise measurement of $\sin(2\beta)$ is made
using $B^0 \to J/\psi K^0$ decays, one might expect that a similar channel
can be used to extract also $\cos(2\beta)$.  
Indeed, methods based on $B^0 \to J/\psi K^0$ decays using the subsequent
evolution of the neutral kaon 
have been proposed~\cite{Azimov:1996fc,Kayser:1997rk,Quinn:2000jy}, 
but appear experimentally challenging.

The first experimental measurement of $\cos(2\beta)$ used the decay mode
$B^0 \to J/\psi K^*(892)$, with $K^*(892) \to K^0_S \pi^0$~\cite{Aubert:2004cp}.
This method relies on the interference between $CP$-even and $CP$-odd
helicity states~\cite{Dunietz:1990cj,Charles:1998gq,Chiang:2000eh}.  A
residual ambiguity due to the unknown sign of the strong phase difference
can be resolved using input from $K\pi$ scattering~\cite{Aston:1987ir} or
from theory~\cite{Suzuki:2001za}.  The most recent
measurements~\cite{Aubert:2004cp,Itoh:2005ks} prefer $\cos(2\beta)>0$ but
with large uncertainty. It will be difficult for LHCb to improve on these
measurements since it is necessary to measure accurately the momentum of the
neutral pion in the final state. 

Other methods using doubly-charmed final states $D^{(*)+}D^{(*)-}K_S$ have
been proposed.  In principle, time-dependent amplitude analyses of these
states would yield information on the weak phase (see the discussion in the
next section).  This is simplest for the decay $B^0 \to
D^+D^-K_S$~\cite{Charles:1998vf,Kayser:1999bt}, where all final state
particles are pseudoscalars so that the Dalitz plot gives a complete
description of the phase space.  The $B$ factory statistics have not yet
enabled this analysis, and due to the high multiplicity of charged tracks in
the final state it may be difficult to study at LHCb.   
If either or both charmed mesons are reconstructed as $D^*$
there are additional degrees of freedom that further complicate the amplitude
analysis.  Fortunately, by integrating over regions of the $B^0 \to
D^{*+}D^{*-}K_S$ phase space, some simplifications are possible, but at the
price of considerable theoretical uncertainty~\cite{Browder:1999ng}.  With
some input from theory, the current experimental
measurements~\cite{Aubert:2006fh,Dalseno:2007hx} prefer $\cos(2\beta)>0$.

\underline{$b \to c \bar{u} d$}

The possibility to measure $\cos(2\beta)$ from $B^0 \to D_{CP} \pi^+\pi^-$
decays~\cite{Charles:1998vf} is the subject of this paper.  A similar analysis
using $B^0 \to D^*_{CP} \pi^+\pi^-$, with $D^*_\pm \to D_\pm \pi^0$ or
$D^*_\pm \to D_\mp \gamma$~\cite{Bondar:2004bi} is possible in principle, but
requires a more involved amplitude analysis including the $D^*$ decay angles.

Another method to measure $\cos(2\beta)$ in $b \to c \bar{u} d$ transitions,
using $B^0 \to Dh^0$ (with $h^0$ being a light neutral meson such as a
$\pi^0$) with time-dependent Dalitz plot analysis of the subsequent neutral
$D$ meson decay to $K_S^0 \pi^+\pi^-$ has been proposed~\cite{Bondar:2005gk}
and implemented~\cite{Krokovny:2006sv,Aubert:2007rp}.  To analyze the latter
decay channel, the $B$ decay vertex position must be determined from the pion
tracks that originate from the $D$ decay.  The lack of primary particles from
the $B$ vertex, together with the necessity to reconstruct the neutral meson,
will make this analysis difficult to carry out in the hadronic environment of
LHCb.

Another interesting approach could be to carry out a simultaneous analysis
of the $B$ and $D$ meson decay Dalitz plots in the $B^0 \to D \pi^+\pi^-$,
$D \to K_S^0 \pi^+\pi^-$ decay chain, thereby combining the method of
Ref.~\cite{Charles:1998vf} and this paper with that of
Ref.~\cite{Bondar:2004bi}. 
Large statistics could be available for such an analysis, since the relevant
branching fractions for both $B$ and $D$ decays are reasonably high.
Although the complete four-dimensional amplitude analysis would be quite
complicated, it may be possible to select regions of the phase space where
simplifications are possible (for example, selecting the $D\rho^0$ dominated
region of the $B$ decay phase space).
In order to reach high precision, however, a complete modelling of the
amplitude is likely to be necessary.

\underline{$b \to c \bar{c} d$}

Several possibilities to measure $\cos(2\beta)$ from $b \to c \bar{c} d$
transitions have been discussed in the literature, but all are experimentally
challenging and none have yet been implemented.  
For example, the interference between $B^0 \to D^{**+}D^-$ and 
$B^0 \to D^+D^{**-}$ decays could be measured in a time-dependent Dalitz plot
analysis of $B^0 \to D^+D^-\pi^0$~\cite{Grossman:1997gd,Charles:1998vf}, or
the interference between $CP$-even and $CP$-odd helicity states could be
probed in a time-dependent analysis of 
$B^0 \to D^{*+}D^{*-}$~\cite{Dunietz:1990cj,Chiang:2000eh}.  Both these
techniques require the reconstruction of a high multiplicity final state as
well as precise understanding of potential misreconstruction effects.
Another interesting possibility is provided by the vector-vector decay 
$B^0 \to J/\psi\rho^0$~\cite{Chiang:2000eh}, though due to the large natural
width of the $\rho$, a time-dependent analysis of $B^0 \to J/\psi \pi^+\pi^-$
may be necessary to incorporate correctly all interference
effects~\cite{Aubert:2007xw}.  This would be a highly challenging analysis,
though it could potentially be studied at LHCb.

Another interesting possibility that has been proposed involves a study of 
helicity amplitudes in the dibaryon decay 
$B^0 \to \Lambda_c \bar{\Lambda}_c$~\cite{Charles:1998gq}.
However, this decay has not yet been observed~\cite{Uchida:2007gx},
meaning that it will be difficult to accrue sufficient statistics for a
precise analysis.

\underline{$b \to q \bar{q} s$}

\nopagebreak
Since measurements of mixing-induced $CP$ violation phenomena in 
$b \to q \bar{q} s$ transitions provide one of the most interesting approaches
to search for effects of physics beyond the Standard
Model~\cite{Grossman:1996ke,Fleischer:1996bv,London:1997zk,Ciuchini:1997zp},
it is clearly important to be able to probe $\cos(2\beta)$ in these
transitions.  Recently, this has been achieved using time-dependent Dalitz
plot analyses of $B^0 \to K_S^0 K^+K^-$ (containing contributions from 
$\phi K_S^0$ and $f_0 K_S^0$ among others)~\cite{Aubert:2008gv} 
and $B^0 \to K_S^0 \pi^+\pi^-$ (containing contributions from $\rho^0 K_S^0$
and $f_0 K_S^0$ among others)~\cite{Aubert:2007vi,Dalseno:2008ww}.
The approach that has been adopted is to obtain values of $\beta$, rather than 
$\cos(2\beta)$ and $\sin(2\beta)$ separately.
The current results indicate that values of $\beta$ closer to the Standard
Model solution than those with $\cos(2\beta)<0$ are preferred, but much more
precise results are needed.  Although LHCb is expected to make some
improvement on the current measurements, these channels provide one of the
motivations for a very high-luminosity electron-positron flavour
factory~\cite{Bona:2007qt,Gershon:2006mt,Browder:2007gg,Browder:2008em}.

\vspace{1ex}

In Table~\ref{tab:cosReview} we summarise the current status of experimental
measurements of $\cos(2\beta)$.  
We have not included results from time-dependent Dalitz plot analyses of 
$B^0 \to K_S^0 K^+K^-$~\cite{Aubert:2008gv} and 
$B^0 \to K_S^0 \pi^+\pi^-$~\cite{Aubert:2007vi,Dalseno:2008ww} 
where the results have been presented in a different format.  
Discounting the results in $B^0 \to D^{*+}D^{*-}K_S^0$,
which suffer from a large theoretical uncertainty, we see that no measurement
has a precision better than about $0.50$, and therefore additional approaches
are very welcome.

\begin{table}[!htb]
  \begin{center}
    \caption{
      Summary of measurements of $\cos(2\beta)$.
      For more details and world averages, see Ref.~\cite{Barberio:2008fa}.
      For all quoted results, the first uncertainty is statistical and the
      second systematic.
      Note that the parameter measured in analyses of 
      $B^0 \to D^{*+}D^{*-}K_S^0$ marked by $(*)$ is not $\cos(2\beta)$ but 
      $(2J_{s2}/J_0)\cos(2\beta)$, where the combination of hadronic
      parameters $(2J_{s2}/J_0)$ is expected to be positive.
      In the BaBar results on $B^0 \to Dh^0$ with $D \to K_S^0 \pi^+\pi^-$,
      the third uncertainty is due to the $D$ decay model; Belle include
      these effects together with other systematic uncertainties.
      The symbol $h^0$ denotes a light neutral meson such as a $\pi^0$.
      \label{tab:cosReview}
    }
    \vspace{1ex}
    \begin{tabular}{c@{\hspace{3mm}}c@{\hspace{3mm}}c}
      \hline
      Experiment & $\cos(2\beta)$ \\
      \hline
      \multicolumn{2}{c}{$B^0 \to J/\psi K^*(892)$ with $K^*(892) \to K^0_S \pi^0$} \\
      BaBar~\cite{Aubert:2004cp} & $3.32\,^{+0.76}_{-0.96} \pm 0.27$ \\
      Belle~\cite{Itoh:2005ks} & $0.56 \pm 0.79 \pm 0.11$ \\
      \hline
      \multicolumn{2}{c}{$B^0 \to D^{*+}D^{*-}K_S^0$} \\
      BaBar~\cite{Aubert:2006fh} (*) & $0.38 \pm 0.24 \pm 0.05$ \\
      Belle~\cite{Dalseno:2007hx} (*) & $-0.23\,^{+0.43}_{-0.41} \pm 0.13$ \\
      \hline
      \multicolumn{2}{c}{$B^0 \to Dh^0$ with $D \to K_S^0 \pi^+\pi^-$} \\
      BaBar~\cite{Aubert:2007rp} & $0.42 \pm 0.49 \pm 0.09 \pm 0.13$ \\
      Belle~\cite{Krokovny:2006sv} & $1.87\,^{+0.40}_{-0.53}\,^{+0.22}_{-0.32}$ \\
      \hline
    \end{tabular}
  \end{center}
\end{table}

In Table~\ref{tab:sinReview} we summarise the existing measurements of
$\sin(2\beta)$ in $b \to c \bar{u} d$ transitions.
Additional approaches that can improve the precision beyond that achieved in
$B^0 \to D^{(*)}_{CP} h^0$ would be very helpful to test the Standard Model
prediction that the value of $\sin(2\beta)$ measured should be the same as
that obtained from $B^0 \to J/\psi K^0$ decays.

\begin{table}[!htb]
  \begin{center}
    \caption{
      Summary of measurements of $\sin(2\beta)$ in $b \to c \bar{u} d$
      transitions. 
      For more details and world averages, see Ref.~\cite{Barberio:2008fa}.
      For all quoted results, the first uncertainty is statistical and the
      second systematic.
      In the BaBar results on $B^0 \to Dh^0$ with $D \to K_S^0 \pi^+\pi^-$,
      the third uncertainty is due to the $D$ decay model; Belle include
      these effects together with other systematic uncertainties.
      \label{tab:sinReview}
    }
    \vspace{1ex}
    \begin{tabular}{c@{\hspace{3mm}}c@{\hspace{3mm}}c}
      \hline
      Experiment & $\sin(2\beta)$ \\
      \hline
      \multicolumn{2}{c}{$B^0 \to D^{(*)}_{CP} h^0$} \\
      BaBar~\cite{Aubert:2007mn} & $0.56 \pm 0.23 \pm 0.05$ \\
      \hline
      \multicolumn{2}{c}{$B^0 \to Dh^0$ with $D \to K_S^0 \pi^+\pi^-$} \\
      BaBar~\cite{Aubert:2007rp} & $0.29 \pm 0.34 \pm 0.03 \pm 0.05$ \\
      Belle~\cite{Krokovny:2006sv} & $0.78 \pm 0.44 \pm 0.22$ \\      
      \hline
    \end{tabular}
  \end{center}
\end{table}

\clearpage
\section{Method}
\label{sec:method}

Detailed descriptions of the method to extract $CP$ violating phases from $B
\to Dh^0$ (where $h^0$ is a light neutral meson such as $\pi^0$, $\eta$, {\it
  etc.}) can be found
elsewhere~\cite{Bondar:2005gk,Fleischer:2003ai,Fleischer:2003aj}.  (Similar
discussions where $h^0$ is a $K_S^0$ meson, relevant for the extraction of the
angle $\gamma$, can also be found in the
literature~\cite{Atwood:2002vw,Gronau:2004gt}.) Here we provide only an
outline of the method.  To simplify the discussion, we initially treat $B$
decays to charm as being flavour-specific -- {\it ie.} we neglect $b \to u
\bar{c} d$ amplitudes, that are suppressed by a factor of approximately 0.02
compared to the favoured $b \to c \bar{u} d$ amplitudes~\cite{Suprun:2001ms}. 
We consider the effects of suppressed amplitudes at a later stage.  

\begin{figure}[!htb]
  \includegraphics[width=0.49\textwidth]{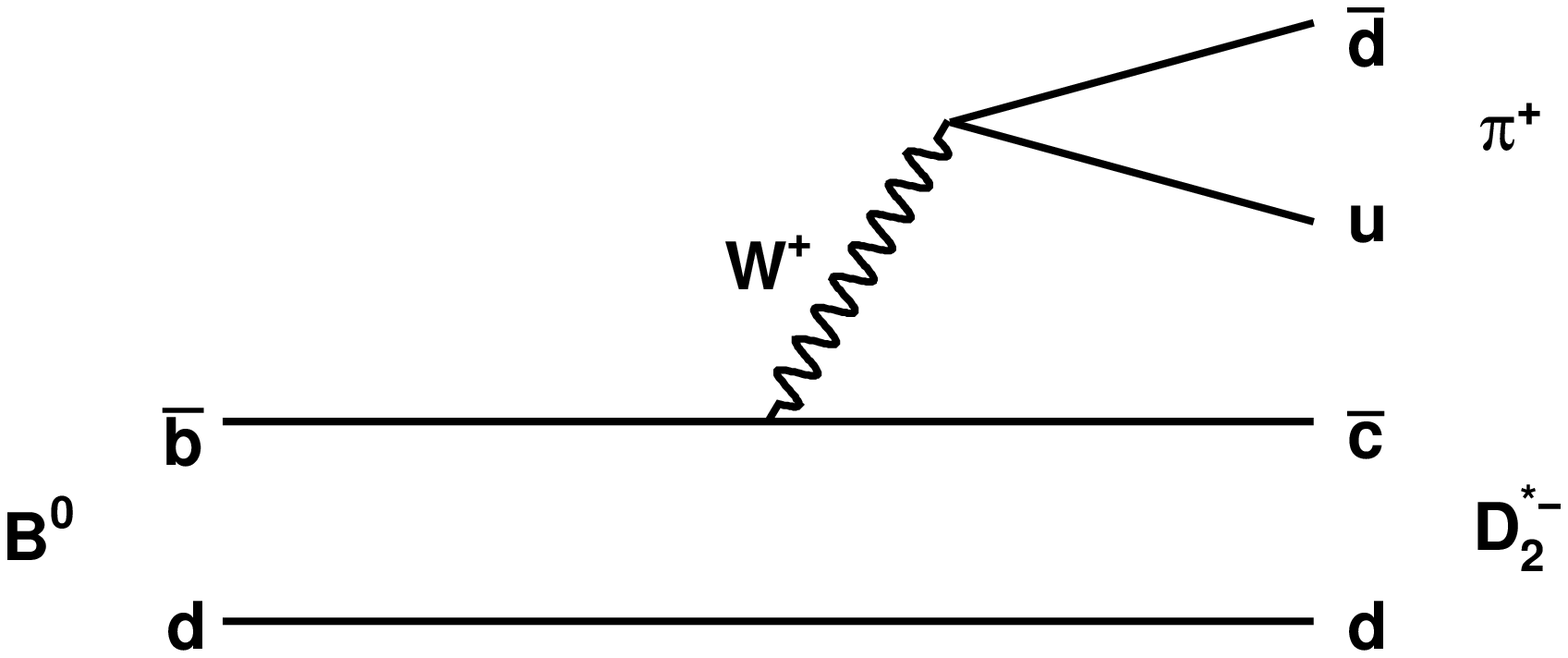}
  \includegraphics[width=0.49\textwidth]{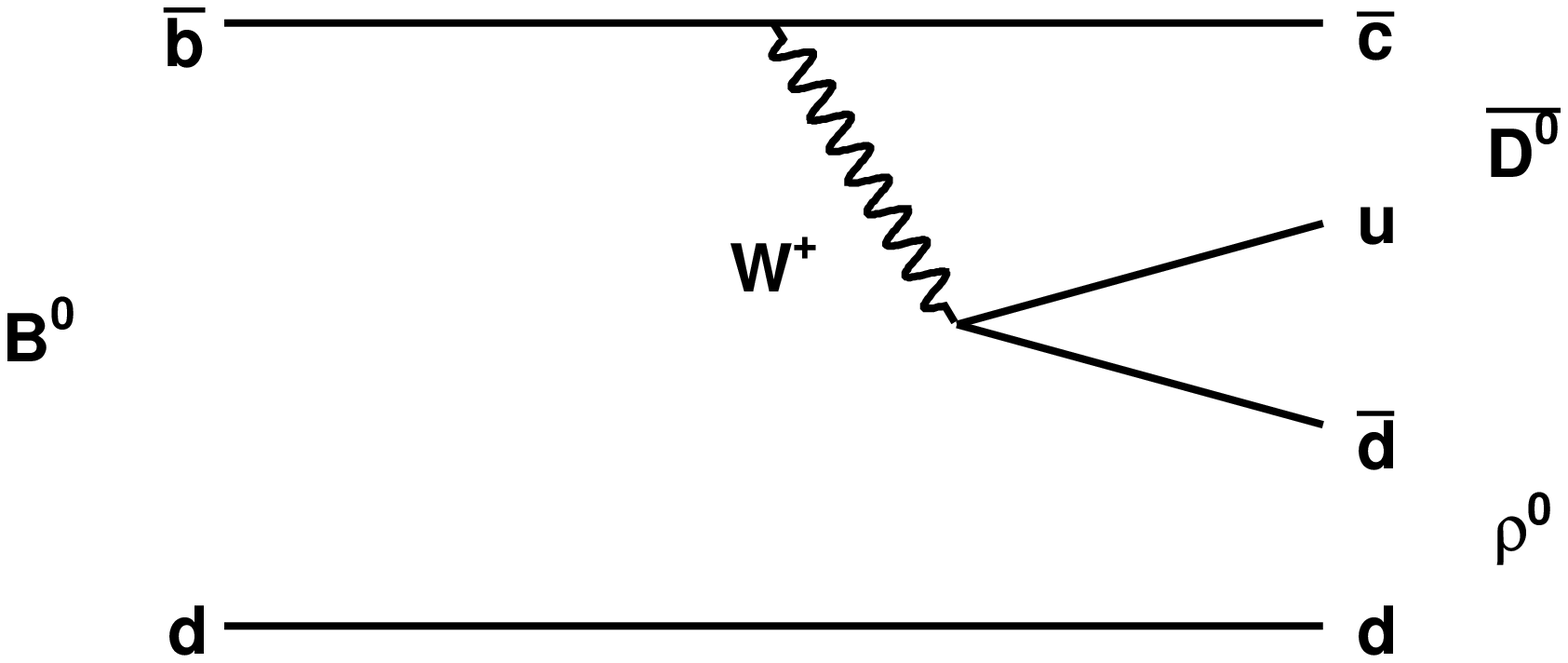}
  \caption{
    Feynman diagrams for 
    (left) a colour-favoured $\bar{b} \to \bar{c} u \bar{d}$
    transition resulting in a $B^0 \to D_2^{*-} \pi^+$ decay,
    and (right) a colour-suppressed $\bar{b} \to \bar{c} u \bar{d}$
    transition resulting in a $B^0 \to \bar{D}^0 \rho^0$ decay.
  }
  \label{fig:feynman}
\end{figure}

Consider the amplitude for a $B^0$ decay to a point in the
$\bar{D}^0\pi^+\pi^-$ Dalitz plot, described by the coordinates $m_+^2 \equiv
m^2(D\pi^+)$ and $m_-^2 \equiv m^2(D\pi^-)$.  We define the amplitude as
\begin{equation}
  A(B^0 \to \bar{D}^0\pi^+\pi^-) \equiv A(m_+^2, m_-^2) = 
  \sum_i c_i F_i(m_+^2, m_-^2) \, ,
\end{equation}
where we have used the isobar formalism to express the amplitude as a sum of
contributions from interfering resonances.  The complex coefficients $c_i$
describe the magnitude and phase of the contribution from each resonance $i$,
while the strong dynamics (lineshapes and angular distributions) are contained
within the $F_i$ functions.  The sum over $i$ will include excited $D$ mesons
({\it eg.} $D_2^{*-}$) and $\pi^+\pi^-$ resonances ({\it eg.} $\rho^0$). 
Feynman diagrams representing these amplitudes are shown in
Fig.~\ref{fig:feynman}.

The amplitude for the $\bar{B}^0$ decay to $D^0 \pi^+\pi^-$ can be written
similarly,
\begin{equation}
  A(\bar{B}^0 \to D^0\pi^+\pi^-) \equiv \bar{A}(m_+^2, m_-^2) = 
  \sum_i \bar{c}_i \bar{F}_i(m_+^2, m_-^2) \, ,
\end{equation}
where the index $i$ is understood to run over the same set of resonances,
except that the charge of the excited $D$ mesons will be opposite compared to
the $B^0$ decay case.  Since $F$ contains strong dynamics only, we can assert
$F_i(m_+^2, m_-^2) = \bar{F}_i(m_-^2, m_+^2)$ and, neglecting direct $CP$ in
$B$ decay, we have $c_i = \bar{c}_i$ both for the excited $D$ mesons and for
the $\pi^+\pi^-$ resonances (which can only be $CP$-even).

Following the usual formalism (see, for
example,~\cite{Harrison:1998yr,Bigi:2000yz,Branco:1999fs}), we write the 
time-dependent decay rate of a $B$ meson that is known to have specific
flavour content (being either $B^0$ or $\bar{B}^0$ at time $\Delta t = 0$) to
a particular final state $f$ or its $CP$ conjugate $\bar{f}$ as 
\begin{eqnarray}
  \Gamma(B^0_{\rm phys} \to f (\Delta t)) & \propto &
  e^{-\left|\Delta t\right|/\tau_{B^0}}
  \left( 
    1 - S_f \sin( \Delta m\Delta t ) + C_f \cos( \Delta m\Delta t ) 
  \right) \, , \\
  \Gamma(\bar{B}^0_{\rm phys} \to f (\Delta t)) & \propto &
  e^{-\left|\Delta t\right|/\tau_{B^0}}
  \left( 
    1 + S_f \sin( \Delta m\Delta t ) - C_f \cos( \Delta m\Delta t )
  \right) \, ,
\end{eqnarray}
where $S_f = 2\,{\rm Im}(\lambda_f)/(1+|\lambda_f^2|)$, 
$C_f = (1-|\lambda_f^2|)/(1+|\lambda_f^2|)$ and $\lambda_f =
\frac{q}{p}\frac{\bar{A}}{A}$.
In the above, $\tau_{B^0}$ is the average lifetime of the neutral $B$ meson,
$\Delta m$ is the mass difference between the two eigenstates of the
$B^0$--$\bar{B}^0$ system which are described in terms of the flavour specific
states by the mixing parameters $q$ and $p$ as
$\left| B_{L\,(H)} \right> = p \left| B^0 \right> +\,(-) \,q \left| \bar{B}^0 \right>$.
(We have neglected the lifetime difference and assumed $CPT$ invariance.)

In the case at hand, $f$ represents a point in the $D\pi^+\pi^-$ Dalitz
plot.  If we consider $D$ decays to $CP$ eigenstates (and neglect direct $CP$
violation in the $D$ system), then we have
\begin{equation}
  \lambda(m_+^2, m_-^2) = 
  \frac{q}{p} \eta_D \frac{\bar{A}(m_+^2, m_-^2)}{A(m_+^2, m_-^2)} \, ,
\end{equation}
where $\eta_D$ is the $CP$ eigenvalue of the $D_{CP}$ state.  With the
further substitution $|q/p| = 1$, ${\rm arg}(q/p) = -2\beta$, both good
approximations in the Standard Model, we can write 
\begin{eqnarray}
  S(m_+^2,m_-^2) = & 
  \frac{
    2 \, {\rm Im} \left( e^{-2i\beta} \eta_D A^* \bar{A} \right)
  }{
    \left|A\right|^2 + \left|\bar{A}\right|^2
  } & =
  \frac{
    2 \, {\rm Im} 
    \left( e^{-2i\beta} \eta_D
      \sum_i c_i^* F_i(m_+^2,m_-^2)^* \sum_j \bar{c}_j \bar{F}_j(m_+^2,m_-^2)
    \right)
  }{
    |\sum_i c_i F_i(m_+^2,m_-^2)|^2 + 
    |\sum_i \bar{c}_i \bar{F}_i(m_+^2,m_-^2)|^2
  } \, , \\
  C(m_+^2,m_-^2) = &
  \frac{ 
    \left|A\right|^2 - \left|\bar{A}\right|^2
  }{
    \left|A\right|^2 + \left|\bar{A}\right|^2
  } & =
  \frac{
    |\sum_i c_i F_i(m_+^2,m_-^2)|^2 - 
    |\sum_i \bar{c}_i \bar{F}_i(m_+^2, m_-^2)|^2
  }{
    |\sum_i c_i F_i(m_+^2,m_-^2)|^2 + 
    |\sum_i \bar{c}_i \bar{F}_i(m_+^2,m_-^2)|^2} \, .
\end{eqnarray}
The numerator of the expression for $S(m_+^2,m_-^2)$ can be written as
$ 2 \eta_D \left( 
  \cos(2\beta) {\rm Im}(A^* \bar{A}) - \sin(2\beta) {\rm Re}(A^* \bar{A})
\right)$,
making  explicit the dependence of mixing-induced $CP$ violation on both 
$\sin(2\beta)$ and $\cos(2\beta)$.  We note that one can choose to fit for
$\sin(2\beta)$ and $\cos(2\beta)$ independently, or alternatively one can fit
directly for $\beta$.  Although the latter appears attractive, the expressions
above make clear that in the former case both parameters appear as
coefficients of physically observable functions of the amplitudes, and thus
one might expect somewhat better statistical behaviour for these observables,
in particular in regions close to physical boundaries.

\begin{figure}[!htb]
  \includegraphics[width=0.49\textwidth]{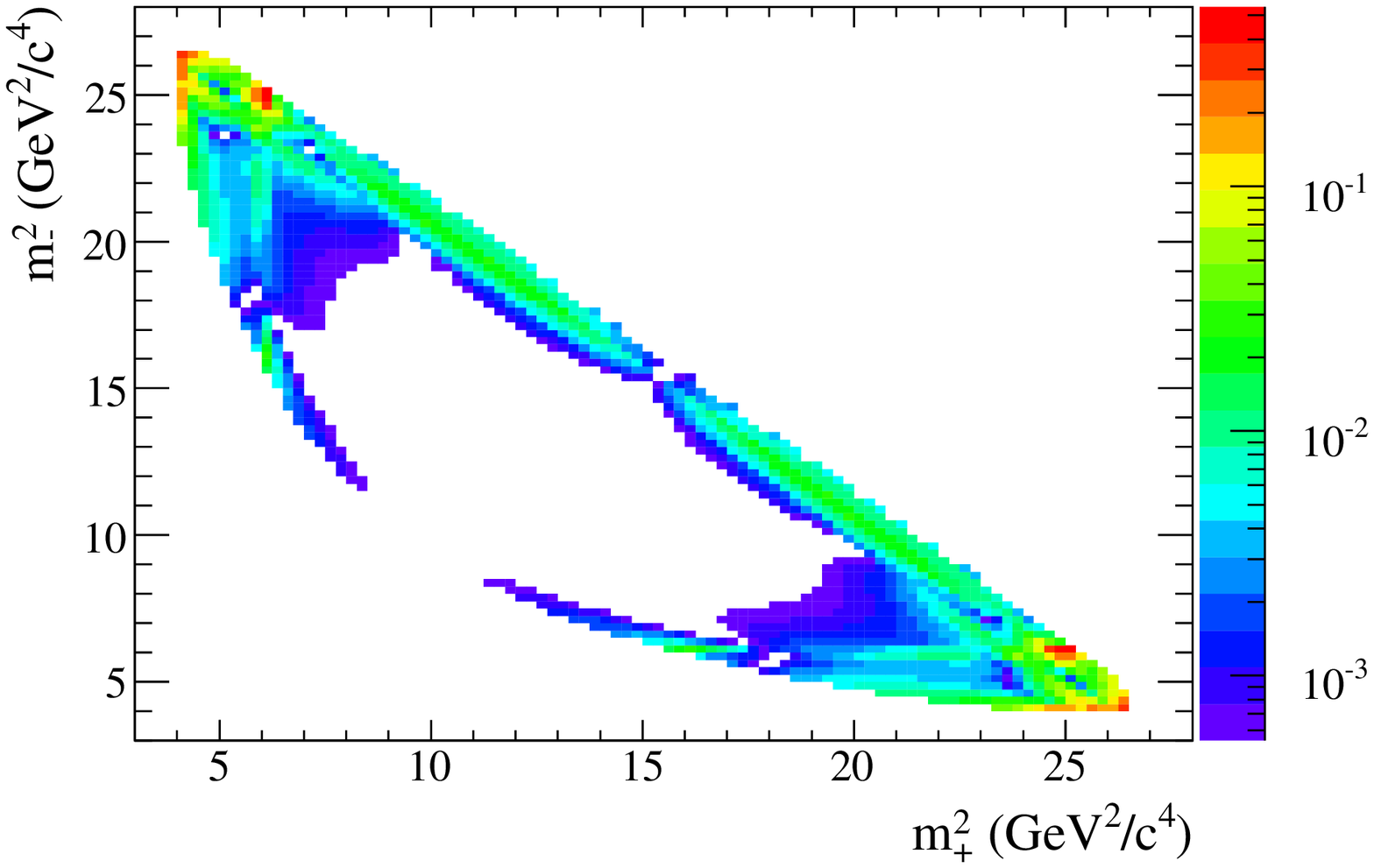}
  \includegraphics[width=0.49\textwidth]{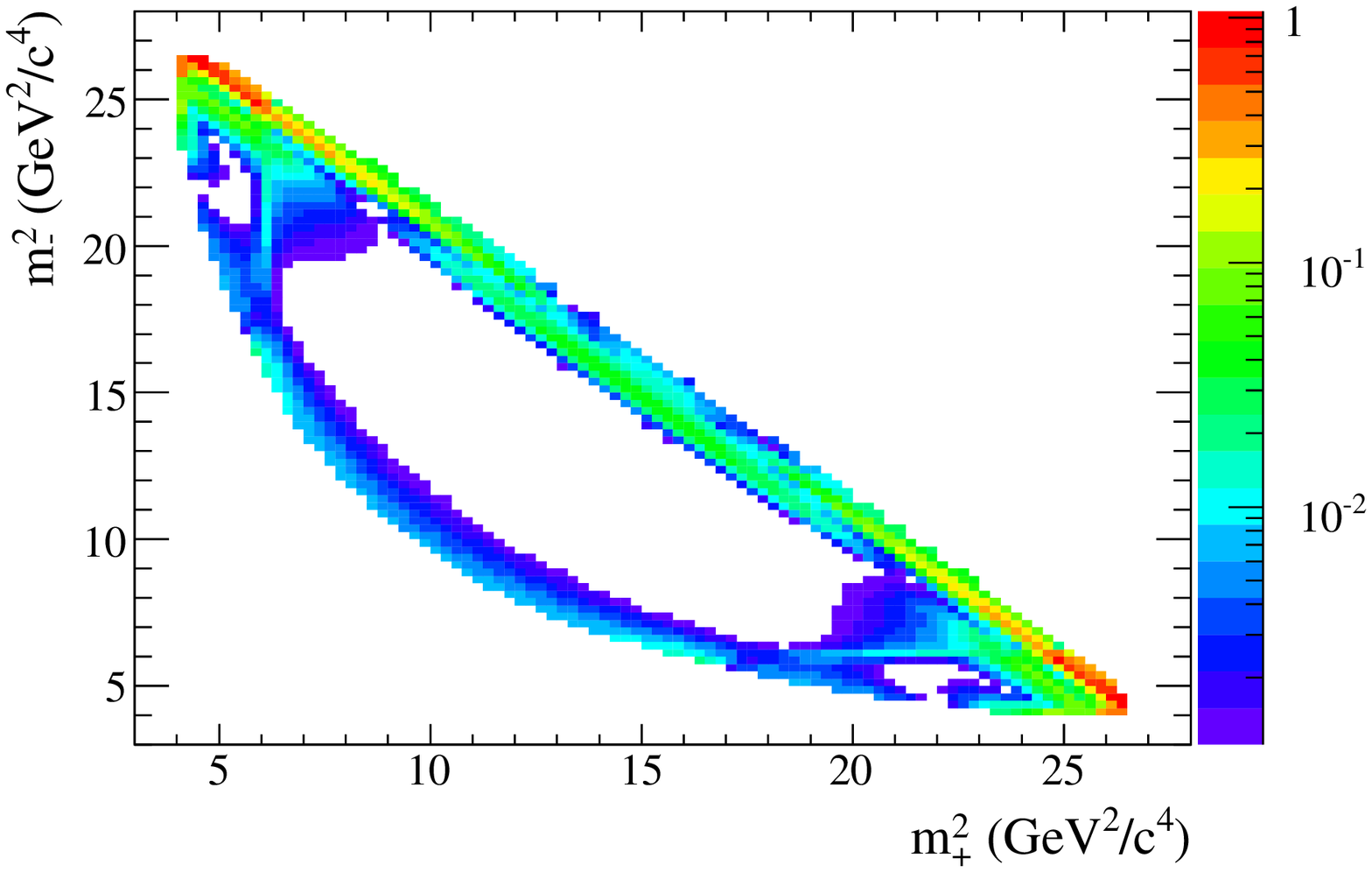}
  \caption{
    Dalitz plot distributions of (left) $\left|{\rm Im}(A^*
    \bar{A})\right|$ and (right) $\left|{\rm Re}(A^* \bar{A})\right|$, 
    which govern the sensitivity to $\cos(2\beta)$ and $\sin(2\beta)$,
    respectively, using our nominal $B^0 \to D_{CP} \pi^+\pi^-$ decay model.
    Note that the $z$-axis is shown on a log scale.
  }
  \label{fig:dp-amplitudes}
\end{figure}

The sensitivity to $\cos(2\beta)$ is proportional to ${\rm Im}(A^* \bar{A})$,
and therefore depends strongly on interference in the Dalitz plot between
resonances with non-trivial phase differences.  To illustrate this point, we
show in Fig.~\ref{fig:dp-amplitudes} the values of ${\rm Im}(A^* \bar{A})$ and
${\rm Re}(A^* \bar{A})$ across the $B^0 \to D_{CP} \pi^+\pi^-$ Dalitz plot,
calculated using our nominal model as described in the next section.
The Dalitz plot is drawn as 
$m_-^2 \equiv m_{D\pi^-}^2$ {\it vs.} $m_+^2 \equiv m_{D\pi^+}^2$.
We emphasise that the Dalitz plot model can be experimentally determined from
the higher statistics $B^0 \to \bar{D}^0 \pi^+\pi^-$ sample with
flavour-specific $\bar{D}^0$ decays.  Consequently, the model uncertainty on
the obtained value of $\cos(2\beta)$ should be controllable.

We conclude this section with a brief discussion of some of the approximations
that have been made in the above formalism.
We have assumed that the $B^0$--$\bar{B}^0$ system contains 
no $CP$ violation in mixing ($|q/p| = 1$), 
negligible lifetime differences ($\Delta \Gamma_d = 0$) 
and no $CPT$ violation.
Furthermore, we have assumed no direct $CP$ violation in $D$ meson decays 
or in $B$ decays to $D\pi^+\pi^-$.
All of these are valid approximations in the Standard Model, and have been 
experimentally tested to good precision.

\begin{figure}[!htb]
  \includegraphics[width=0.49\textwidth]{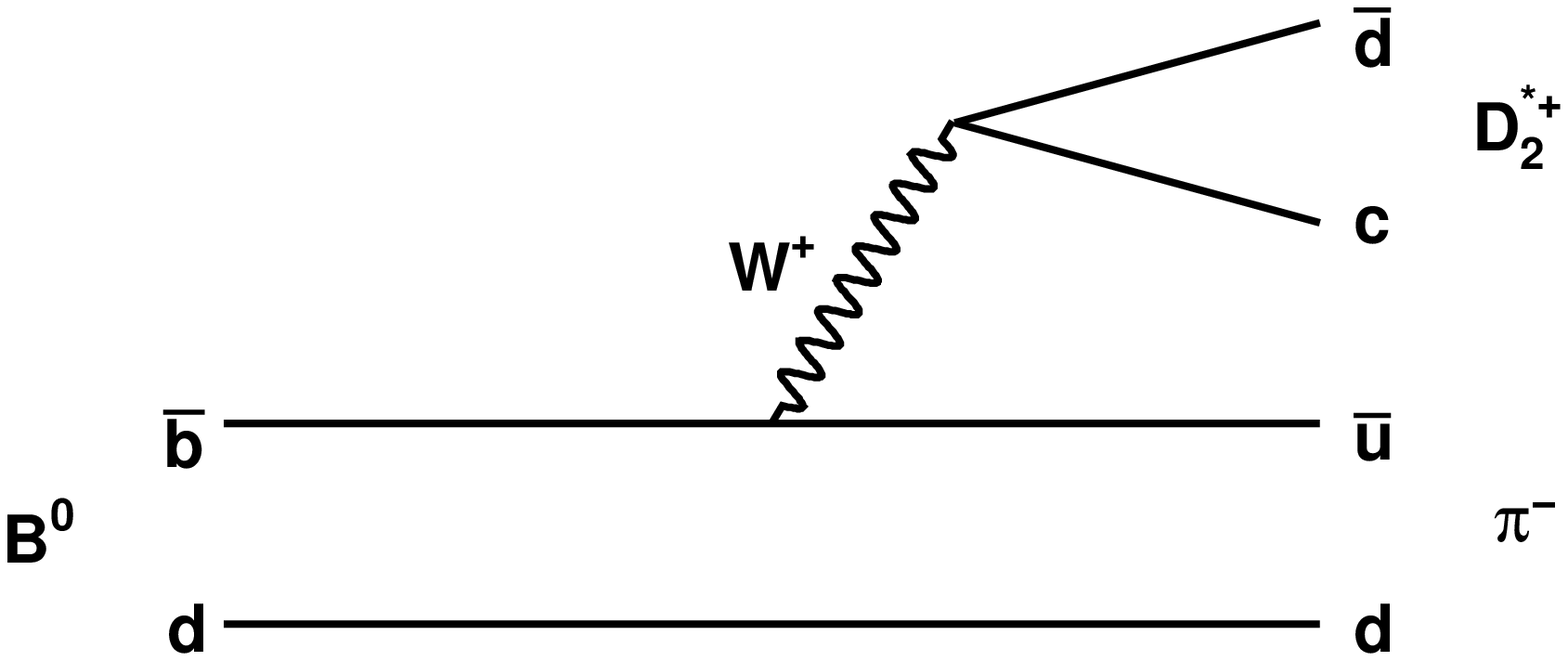}
  \includegraphics[width=0.49\textwidth]{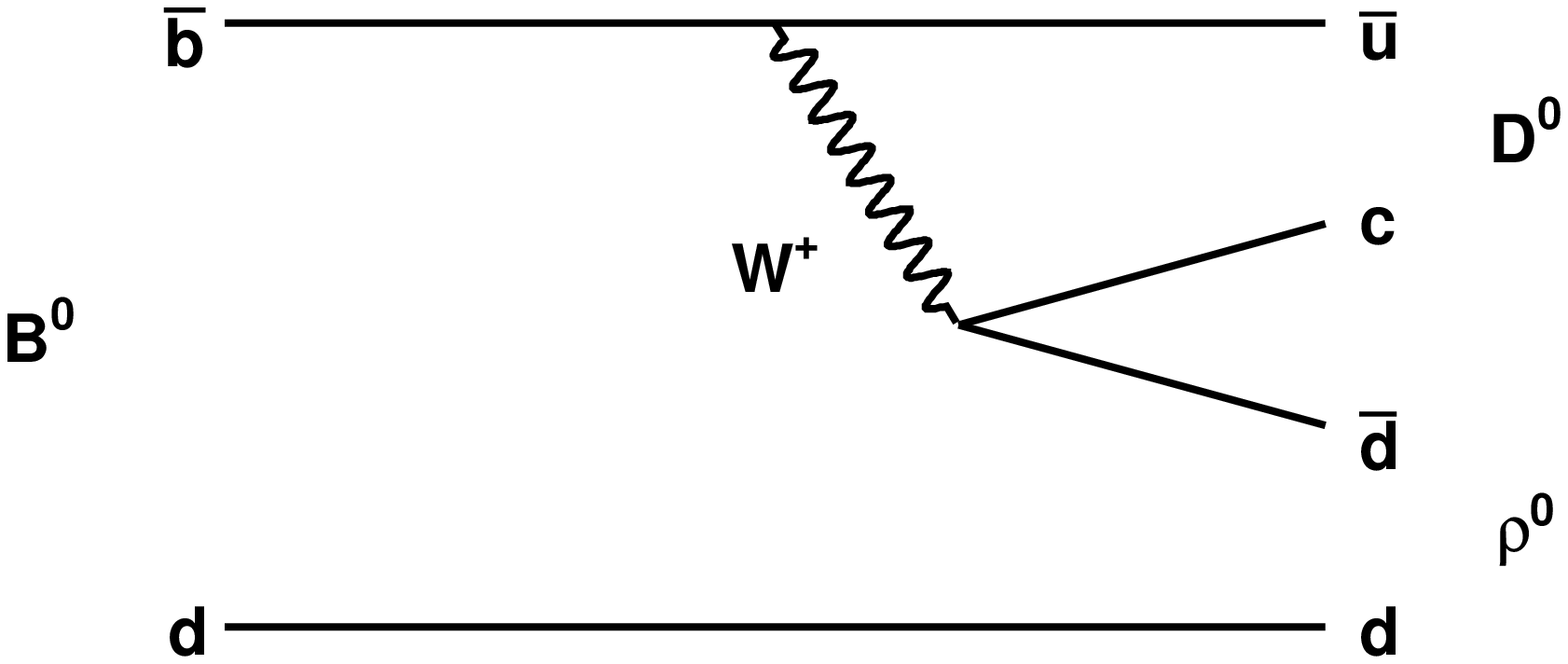}
  \caption{
    Feynman diagrams for Cabibbo-suppressed transitions:
    (left) $\bar{b} \to \bar{u} c \bar{d}$ 
    resulting in $B^0 \to D_2^{*+} \pi^-$,
    and (right) $\bar{b} \to \bar{u} c \bar{d}$
    resulting in $B^0 \to D^0 \rho^0$.
  }
  \label{fig:feynman-sup}
\end{figure}

We have also until now neglected contributions from the suppressed 
$b \to u \bar{c} d$ amplitudes, illustrated in Fig.~\ref{fig:feynman-sup}.
The relative weak phase between $b \to c \bar{u} d$ and $b \to u \bar{c} d$
amplitudes is given by the CKM Unitarity Triangle angle $\gamma$.
If the contribution of the suppressed decays is significant, it leads to
some interesting phenomenology, including the potential to measure
$\sin(2\beta+\gamma)$ from resonant amplitudes such as $D_2^{*\pm}\pi^\mp$,
and to measure $\gamma$ from rates and direct $CP$ violation effects in
modes such as $D\rho^0$.  Both measurements would, however, benefit from the
larger statistics that are available by reconstructing the $D$ meson in a
flavour-specific decay mode and are anyway unlikely to be competitive with
similar measurements using $D^{(*)\pm}\pi^\mp$ and $DK$ final states,
respectively.  This serves to illustrate how flavour-specific $D$ decay
modes can be used to control model uncertainties, as well as other
experimental systematic uncertainties, in the analysis.  A further corollary
is that, unless the suppressed amplitudes are accounted for, there can be
small biases on the extracted values of $2\beta$ that are measured from 
$B^0 \to D_{CP}\pi^+\pi^-$, with the biases opposite in sign for $D$ mesons
reconstructed in $CP$-even and $CP$-odd final states~\cite{Fleischer:2003aj}.
The suppressed amplitudes therefore introduce some model dependence into the
results, which can be tested by adding suppressed (``wrong-sign'') $D_2^*$
resonances (for example) into the model, by relaxing the constraint 
$c_i = \bar{c}_i$, and by checking the consistency of results of independent
fits to the samples with $CP$-even and $CP$-odd $D$ decays.

\clearpage
\section{Feasibility Study}
\label{sec:feasibility}

\subsection{Nominal Model}
\label{subsec:nominal}

We form a nominal $B^0 \to \bar{D}^0 \pi^+\pi^-$ Dalitz plot model based on
results from the Belle Collaboration~\cite{Kuzmin:2006mw}.
The most prominent contributions to the Dalitz plot are found to be 
$B^0 \to D_2^{*-}\pi^+\, , \, D_2^{*-} \to \bar{D}^0 \pi^-$;
$B^0 \to D_0^{*-}\pi^+\, , \, D_0^{*-} \to \bar{D}^0 \pi^-$;
$B^0 \to D_v^{*-}\pi^+\, , \, D_v^{*-} \to \bar{D}^0 \pi^-$;
$B^0 \to \bar{D}^0 \rho^0 \, , \, \rho^0 \to \pi^+\pi^-$ and 
$B^0 \to \bar{D}^0 f_2 \, , \, f_2 \to \pi^+\pi^-$.
The parameters of these resonances are taken from the Particle Data Group
(PDG)~\cite{Amsler:2008zz} and are summarised in
Table~\ref{tab:resonanceParams}.
The particle denoted $D_v^{*-}$ is a virtual $D^{*-}$ meson.  
The $D^{*-}$ itself is too long-lived to cause interference and, following
Belle~\cite{Kuzmin:2006mw}, we veto the region dominated by this contribution
by excluding the invariant mass range 
$2.00 \, {\rm GeV}/c^2 < m_{D\pi} < 2.02 \, {\rm GeV}/c^2$ from the analysis.

\begin{table}[!htb]
  \begin{center}
    \caption{
      Parameters of the resonances used in our nominal model.
      These values are taken from the PDG~\cite{Amsler:2008zz}.
      \label{tab:resonanceParams}
    }
    \vspace{1ex}
    \begin{tabular}{c@{\hspace{3mm}}c@{\hspace{3mm}}c}
      \hline
      Resonance & Mass (MeV/$c^2$) & Width (MeV) \\
      \hline
      $D_2^{*-}$ & $2461.1 \pm 1.6$ & $43 \pm 4$ \\
      $D_0^{*-}$ & $2352 \pm 50$    & $261 \pm 50$ \\
      $D_v^{*-}$ & $2010.27 \pm 0.17$ & $0.096 \pm 0.022$ \\
      $\rho^0$   & $775.49 \pm 0.34$ & $149.4 \pm 1.0$  \\
      $f_2$      & $1275.1 \pm 1.2$ & $185.0\,^{+2.9}_{-2.4}$ \\
      \hline
    \end{tabular}
  \end{center}
\end{table}

The model described above is sufficient for the purposes of our feasibility
study, although the $\bar{D}^0 \pi^+\pi^-$ decay amplitude may include
additional resonant or nonresonant terms.  Belle found a possible contribution
from $B^0 \to \bar{D}^0 f_0(600)$ (the $f_0(600)$ is sometimes known as the
$\sigma$ meson).  Ultimately the decay model and its uncertainty should be
determined from $B^0 \to \bar{D}^0 \pi^+\pi^-$ with flavour-specific 
$\bar{D}^0$ decays as a part of the analysis.

\subsection{Event Generation}
\label{subsec:generation}

We estimate the number of $B^0 \to D_{CP} \pi^+\pi^-$ events that we expect in
the final BaBar dataset as follows.
Belle found $2909 \pm 115$ $B^0 \to \bar{D}^0 \pi^+\pi^-\,, \ \bar{D}^0 \to
K^+\pi^-$ events in a data sample of $388 \times 10^{6}$ $B\bar{B}$ pairs.
We assume similar selection efficiency to that in the Belle analysis, and
scale according to the size of the final BaBar dataset: $467 \times 10^{6}$
$B\bar{B}$ pairs. 
We calculate the expected numbers of events in $CP$-even ($K^+K^-$ and
$\pi^+\pi^-$) and $CP$-odd ($K_S^0\pi^0$ and $K_S^0\omega$) events using
scaling factors calculated from event yields in a recent analysis of 
$B^+ \to DK^+$ from BaBar~\cite{Aubert:2008yk}.
These factors are $(8.6\pm0.2)\%$ for $K^+K^-$ and $(3.1\pm0.1)\%$ for 
$\pi^+\pi^-$, giving a total $(11.7\pm0.2)\%$ for $CP$-even;
correspondingly $(8.9\pm0.2)\%$ for $K_S^0\pi^0$ and $(3.3\pm0.1)\%$ for 
$K_S^0\omega$ sum to a total $(12.2\pm0.2)\%$ for $CP$-odd.
Taking all these factors into account, we expect approximately 410 $CP$-even
events and 430 $CP$-odd events in the final BaBar data sample.  

\begin{figure}[!htb]
  \includegraphics[width=0.75\textwidth]{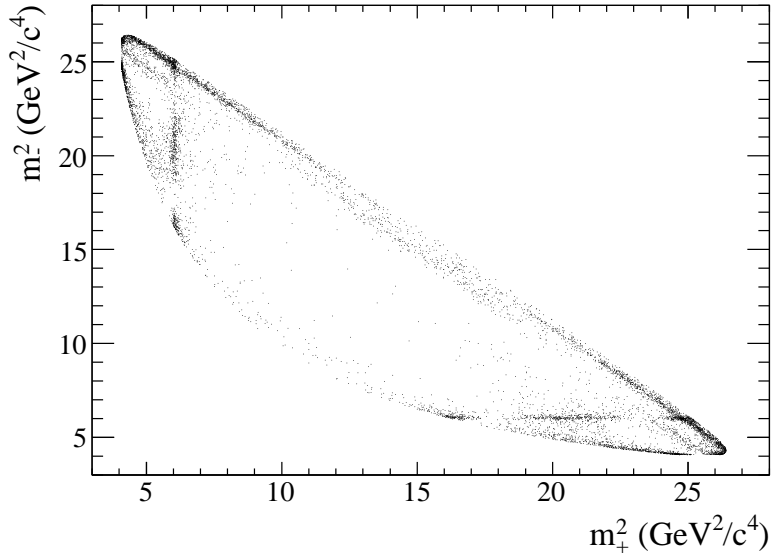}
  \caption{
    Dalitz plot distribution of the generated events using our nominal
    model.  The statistics shown here are ten times that expected in the
    full BaBar dataset.
  }
  \label{fig:dp-dist}
\end{figure}

\begin{figure}[!htb]
  \includegraphics[width=0.48\textwidth]{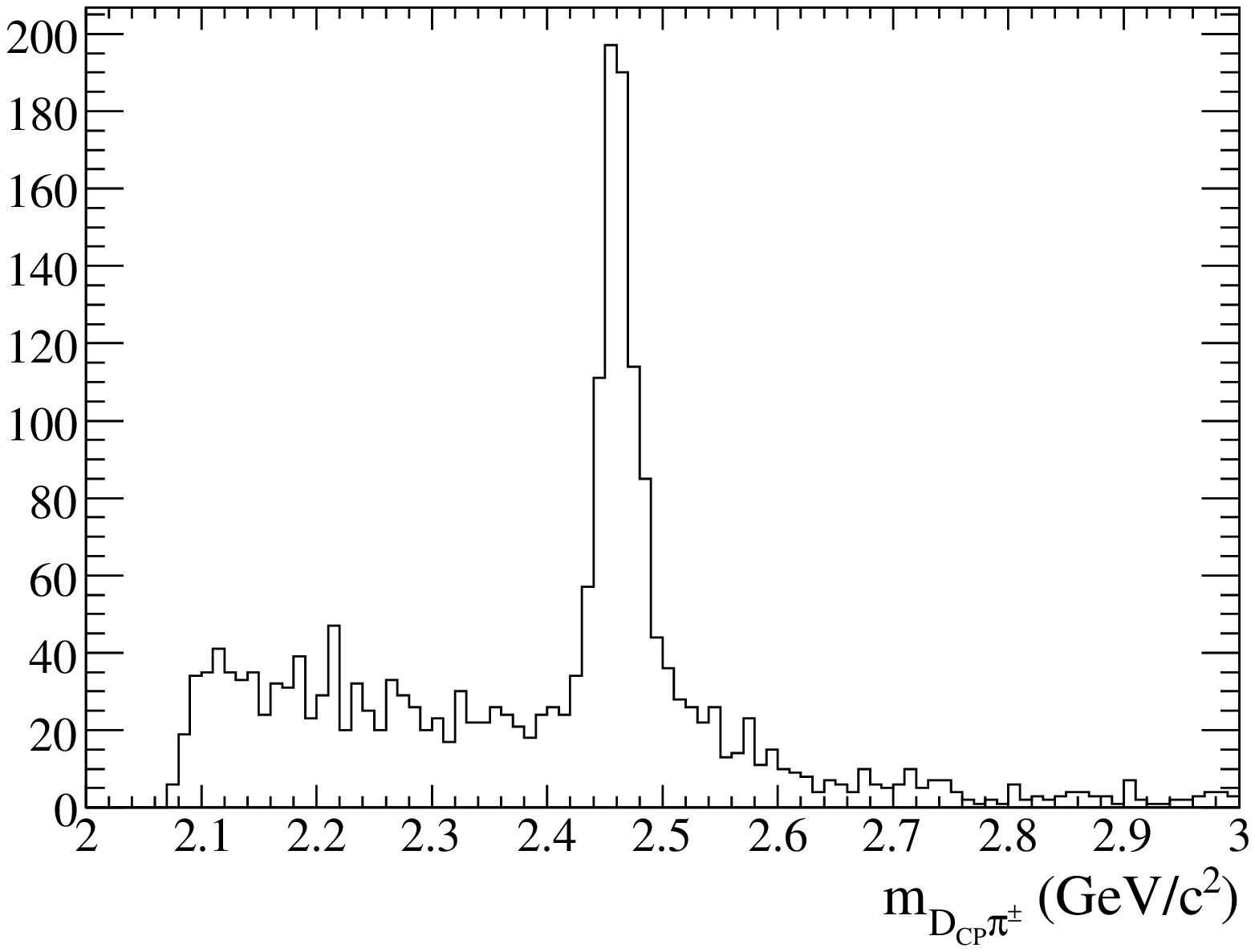}
  \includegraphics[width=0.48\textwidth]{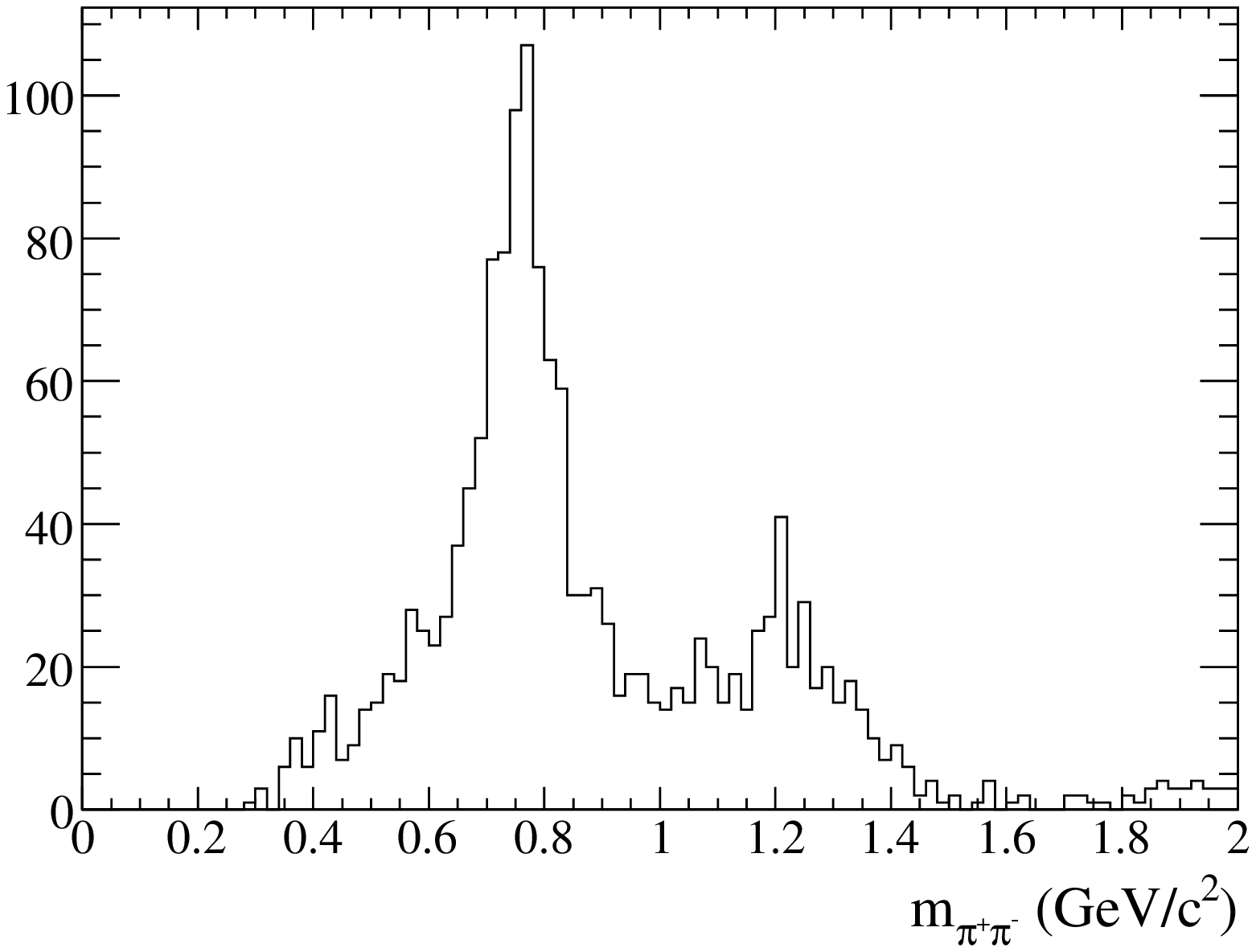}
  \caption{
    Projections onto (left) $m_{D_{CP}\pi^\pm}$ and (right) $m_{\pi^+\pi^-}$
    of the generated events using our nominal model.  The statistics shown
    here are ten times that expected in the full BaBar dataset. 
    To avoid artefacts due to reflections, in the $m_{D_{CP}\pi^\pm}$
    projection we require $m_{\pi^+\pi^-} > 2.0 \ {\rm GeV}/c^2$; in the
    $m_{\pi^+\pi^-}$ projection we require 
    $m_{D_{CP}\pi^\pm} > 2.75 \ {\rm GeV}/c^2$ (both combinations).
    Structures due to (left) $D_2^{*-}$ and (right) $\rho^0$ and $f_2$
    resonances are clearly apparent.
  }
  \label{fig:dp-projections}
\end{figure}

We use the {\tt Laura++} package to generate and fit events. 
This package has been largely developed by the authors and used in several
recent analyses published by the BaBar
Collaboration~\cite{Aubert:2005sk,Aubert:2005ce,Aubert:2007xb,Aubert:2008bj,Aubert:2008rr}.
We generate events using $c_i$ parameters based on results of the Belle
analysis of $B^0 \to \bar{D}^0 \pi^+\pi^-$~\cite{Kuzmin:2006mw}
and $c_i = \bar{c}_i$, using a convention in which all $F_i$ functions are
normalised to unity when integrated over the Dalitz plot.  
We assume a reconstruction efficiency that does not vary across the
Dalitz plot and neglect misreconstruction effects.
Initially, we also neglect backgrounds from other $B$ decays or other
processes, but we include these events in the study at a later stage,
discussed below. We simulate effects due to $\Delta t$ resolution and
misidentification of the flavour of the tagging $B$ meson using standard
resolution functions and parameters from BaBar~\cite{Aubert:2004zt}.
In the fit we float the real and imaginary parts of all $c_i$ except those for
the $D_2^{*-}$ which are fixed as reference parameters.  
We also float both $\cos(\phi_{\rm mix})$ and $\sin(\phi_{\rm mix})$, 
where $\phi_{\rm mix} = 2\beta$ in the Standard Model.
The generated distribution of events in the Dalitz plot for ten times the
expected statistics can be seen in Fig.~\ref{fig:dp-dist}.
Projections onto $m_{D_{CP}\pi^\pm}$ and $m_{\pi^+\pi^-}$ can be seen in
Fig.~\ref{fig:dp-projections}.

\subsection{Results}
\label{subsec:results}

\begin{figure}[!htb]
  \includegraphics[width=0.49\textwidth]{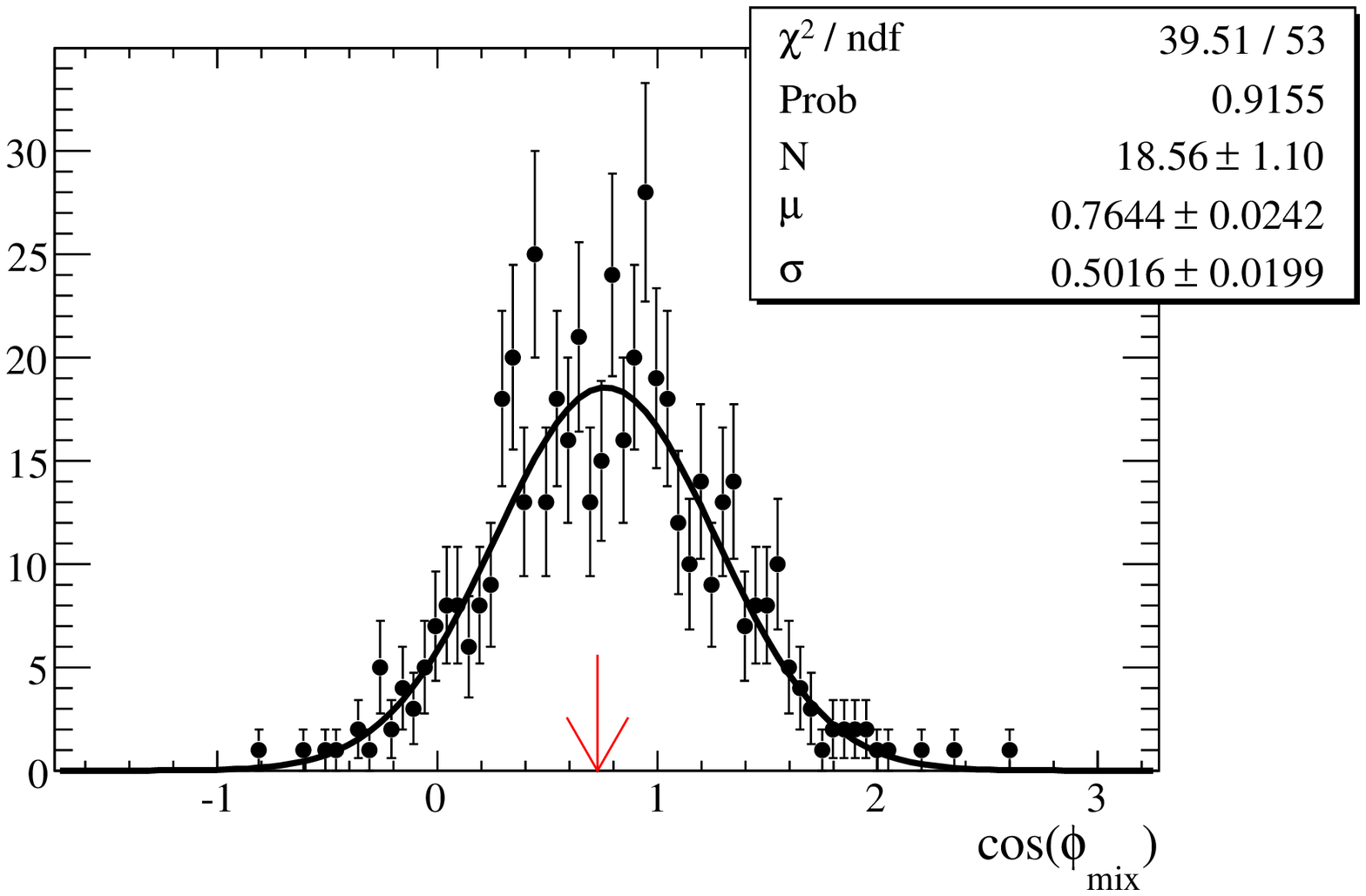}
  \includegraphics[width=0.49\textwidth]{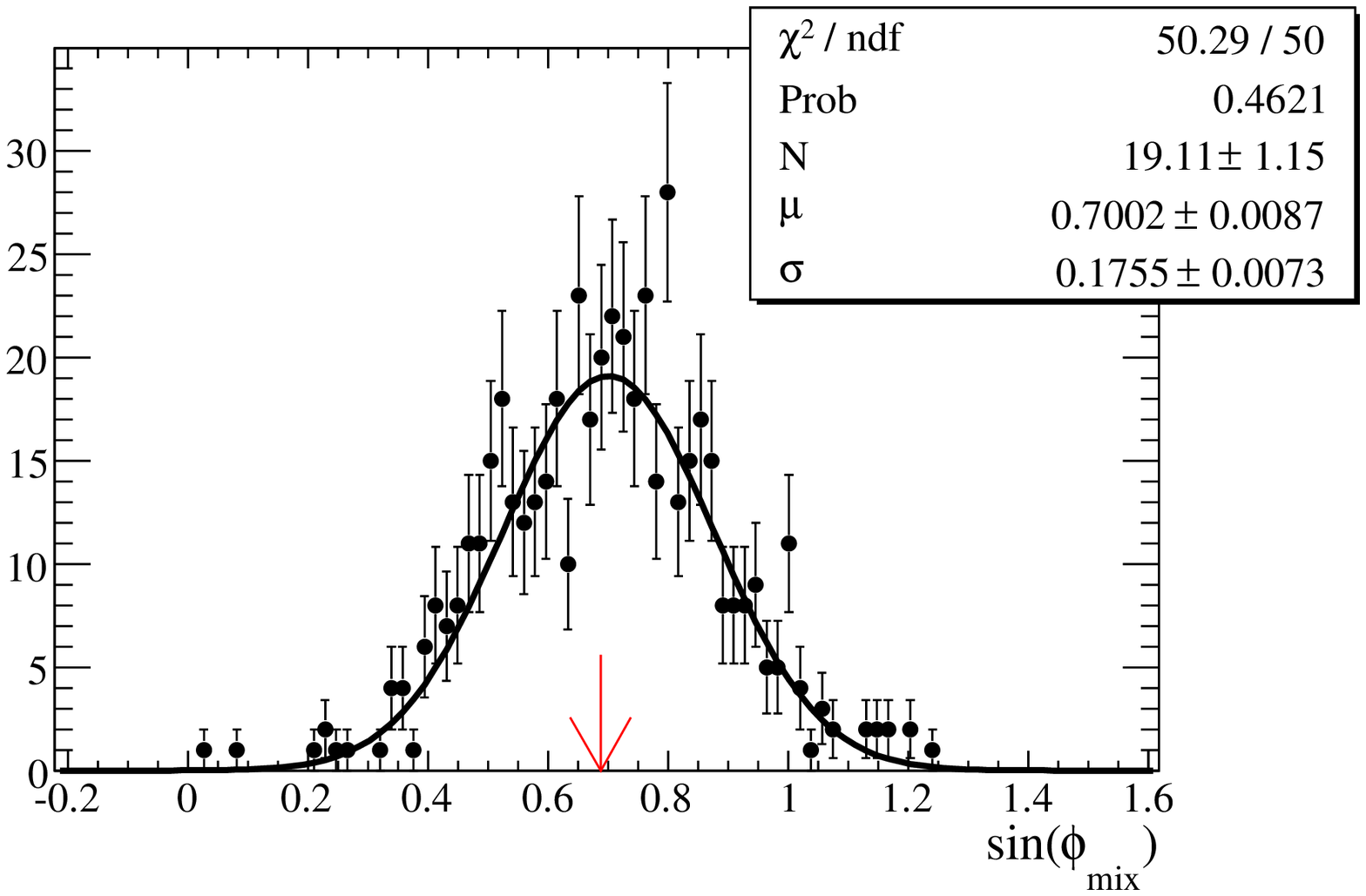}
  \caption{
    Distributions of the fitted values of (left) $\cos(\phi_{\rm mix})$ and
    (right) $\sin(\phi_{\rm mix})$ from the pure signal study.
    The resolutions are $0.50\pm0.02$ and $0.18\pm0.01$, respectively.
    The red arrows indicate the generated values, which are based on the
    value $2\beta = 21.7^\circ$.
  }
  \label{fig:sigOnly}
\end{figure}

The distributions of the fitted results for $\cos(\phi_{\rm mix})$ and
$\sin(\phi_{\rm mix})$ from 500 pseudo-experiments generated as described in
the previous subsection are shown in Fig.~\ref{fig:sigOnly}.
We are clearly able to determine both $\cos(\phi_{\rm mix})$ and
$\sin(\phi_{\rm mix})$, with spreads of the distributions of about $0.50$ and
$0.18$ respectively.  (For comparison, the means of the distributions of the
uncertainties on these parameters that result from the fits are found to be
$0.43$ and $0.17$ respectively, confirming the approximately Gaussian
nature of these parameters that is also evident in Fig.~\ref{fig:sigOnly}.)
Small biases in the fit results are found to disappear when the number of
events per experiment is increased.
All fitted $c_i$ parameters are found to be similarly consistent with the
input values.

\begin{figure}[!htb]
  \includegraphics[width=0.49\textwidth]{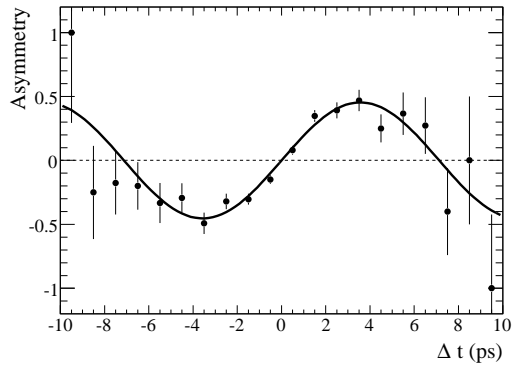}
  \caption{
    Asymmetry between $B^0$ and $\bar{B^0}$ tags as a function of
    $\Delta t$ in the region of the $\rho^0$ resonance 
    ($0.70 \, {\rm GeV}/c^2 < m_{\pi^+\pi^-} < 0.85 \, {\rm GeV}/c^2$).  
    Only events in the best tagging category are shown.
    The statistics shown here are 500 times that expected in the full BaBar
    dataset.
  }
  \label{fig:deltaT}
\end{figure}

Further to illustrate the method, we show in Fig.~\ref{fig:deltaT} the
$\Delta t$ asymmetry for events in the region of the $\rho^0$ resonance in the
$m_{\pi^+\pi^-}$ distribution.
The asymmetry is between events where the other $B$ meson has been
identified (``tagged'') as a $B^0$ or a $\bar{B}^0$ 
by the charge of the lepton produced through its semileptonic decay.
If this region contained $B^0 \to D_{CP} \rho^0$ decays alone, the asymmetry
would give $\sin(2\beta)\sin(\Delta m \Delta t)$, smeared by experimental
effects ($\Delta t$ resolution and misidentification of the flavour of the
tagging $B$ meson).
The plot contains only events containing a $D$ meson reconstructed in a
$CP$-even decay mode and has 500 times the statistics expected from the
final BaBar dataset.

\begin{figure}[!htb]
  \includegraphics[width=0.49\textwidth]{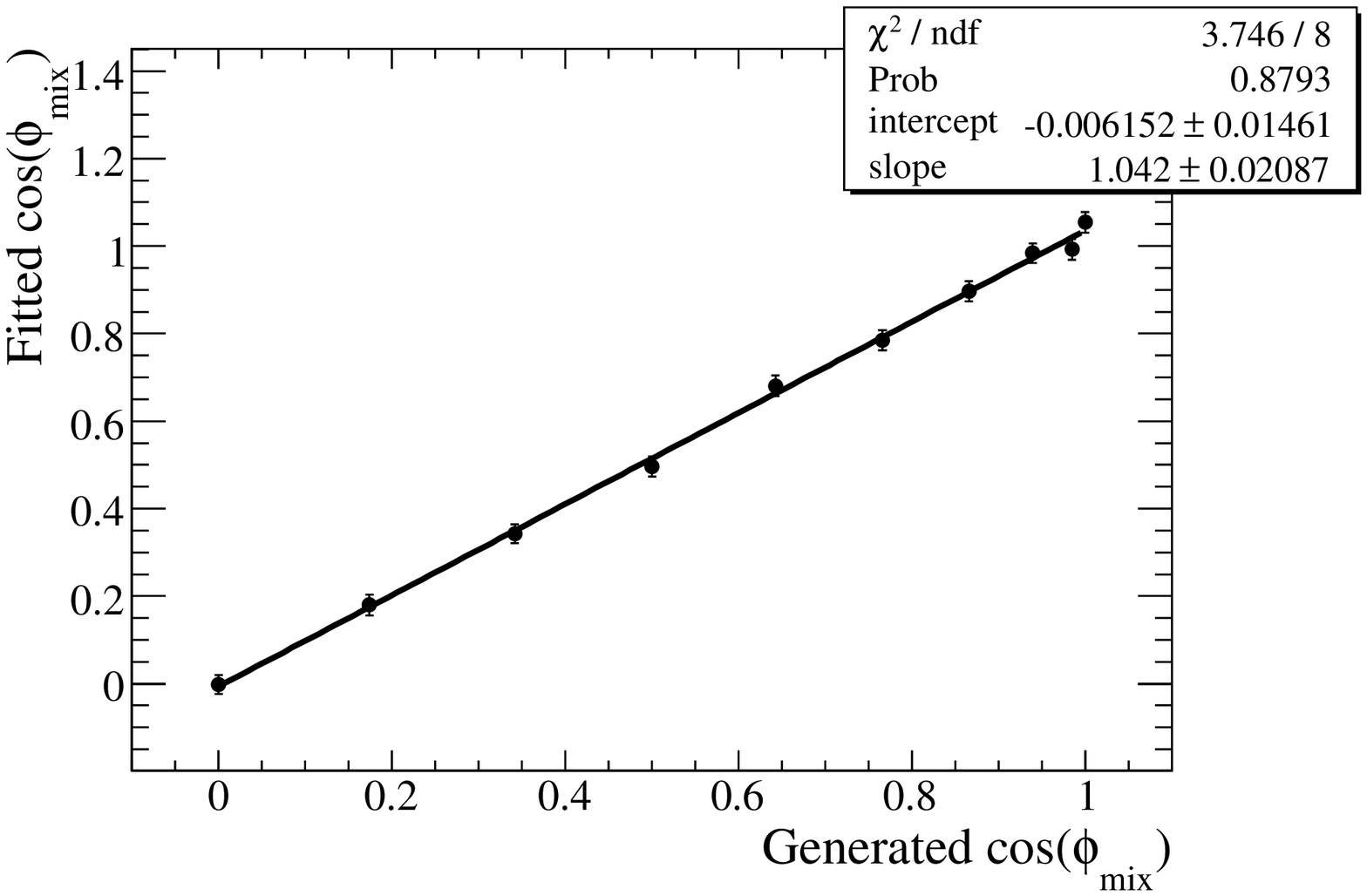}
  \includegraphics[width=0.49\textwidth]{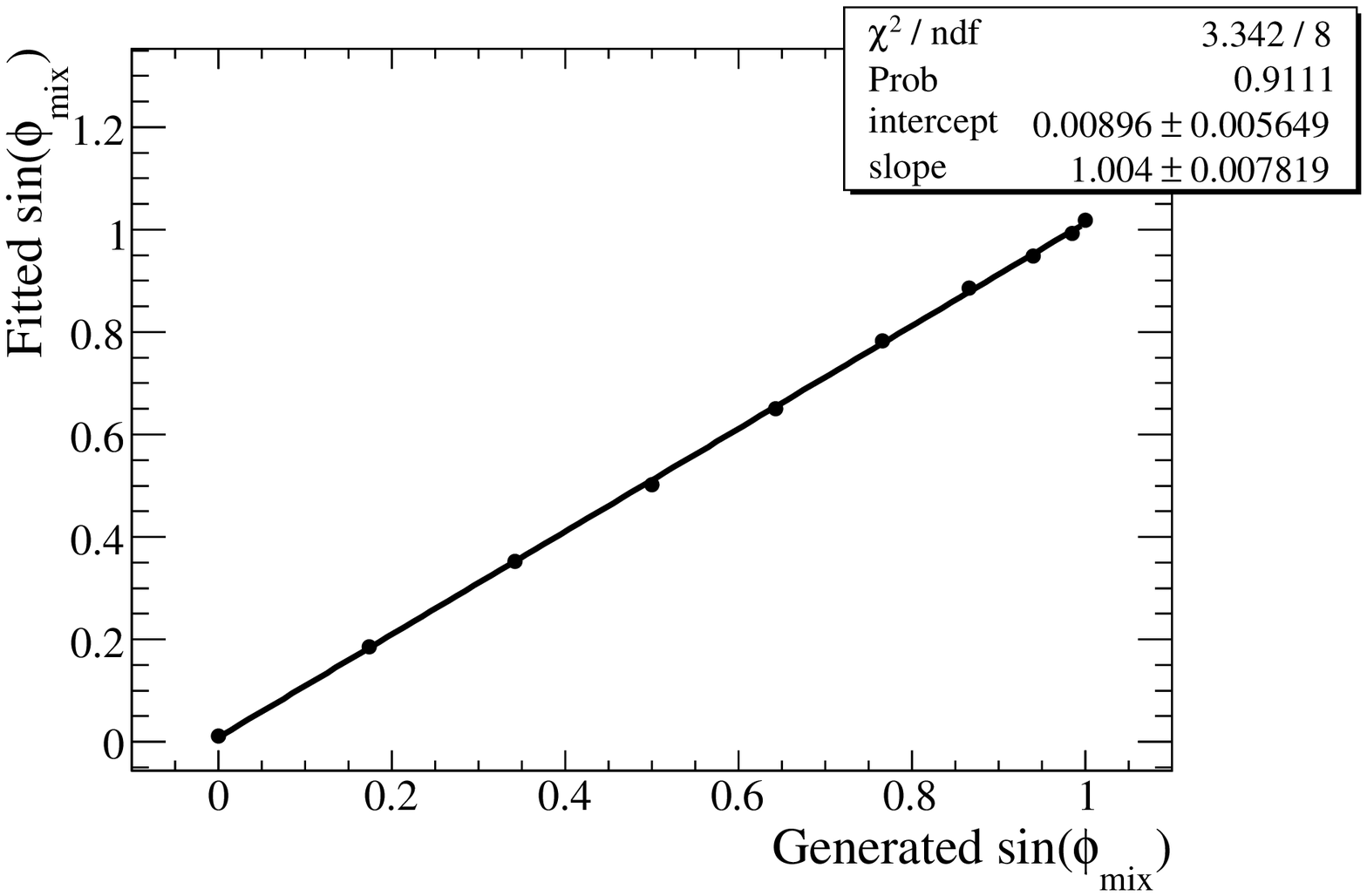}
  \caption{
    Fitted values as a function of the generated values of 
    (left) $\cos(\phi_{\rm mix})$ and (right) $\sin(\phi_{\rm mix})$.
    In the latter plot the error bars are too small to be seen.
  }
  \label{fig:linearity}
\end{figure}

To ensure that our results do not depend on the true value of 
$\phi_{\rm mix}$ we repeat the process with a number of different input
values.  As shown in Fig.~\ref{fig:linearity}, we determine the correct
values of $\cos(\phi_{\rm mix})$ and $\sin(\phi_{\rm mix})$ in all cases.
We also find that uncertainties on these parameters do not depend strongly
on the input values -- a variation of 4\% (18\%) is found in the mean
uncertainty on the fitted value of $\cos(\phi_{\rm mix})$ 
($\sin(\phi_{\rm mix})$), with the uncertainty being largest when 
$\phi_{\rm mix}$ is smallest.

\subsection{Experimental Complications}
\label{subsec:expt}

In a true experimental environment there will be events from background
processes that will have a specific structure in the Dalitz plot and also
in $\Delta t$.  The presence of these events will complicate the analysis
and could degrade the sensitivity to the parameters of interest.  In order
to attempt to estimate the scale of these effects we repeat the study
including background events.
The Belle analysis~\cite{Kuzmin:2006mw} indicates that the level of
background is approximately the same as that of signal once they apply a
selection on the discriminating kinematic variables 
$m_{\rm ES}$ and $\Delta E$ (for definitions of these variables, see for
example~\cite{Aubert:2008bj}).  
As such we include the same number of background events as signal in our
samples.  We take the distribution of the background events in the Dalitz
plot from the Belle paper and use a delta function for the true $\Delta t$
distribution.
In order to provide further discrimination between signal and background we
also include $m_{\rm ES}$ and $\Delta E$ in the fit.  We use Gaussian
shapes to describe the signal distribution of both of these variables,
whilst for background we use the ARGUS shape~\cite{Albrecht:1990cs} and a
linear function, respectively.

\begin{figure}[!htb]
  \includegraphics[width=0.49\textwidth]{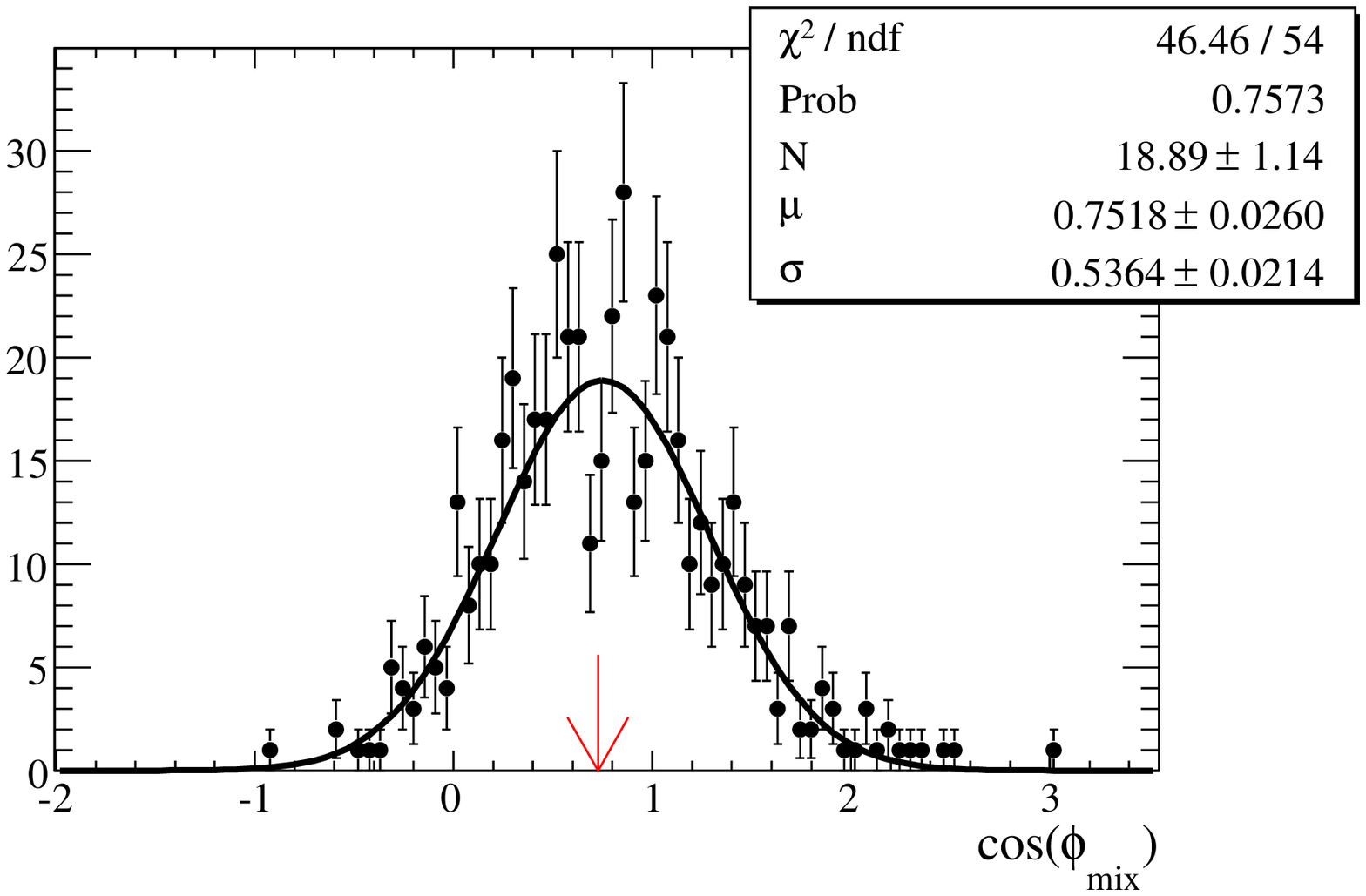}
  \includegraphics[width=0.49\textwidth]{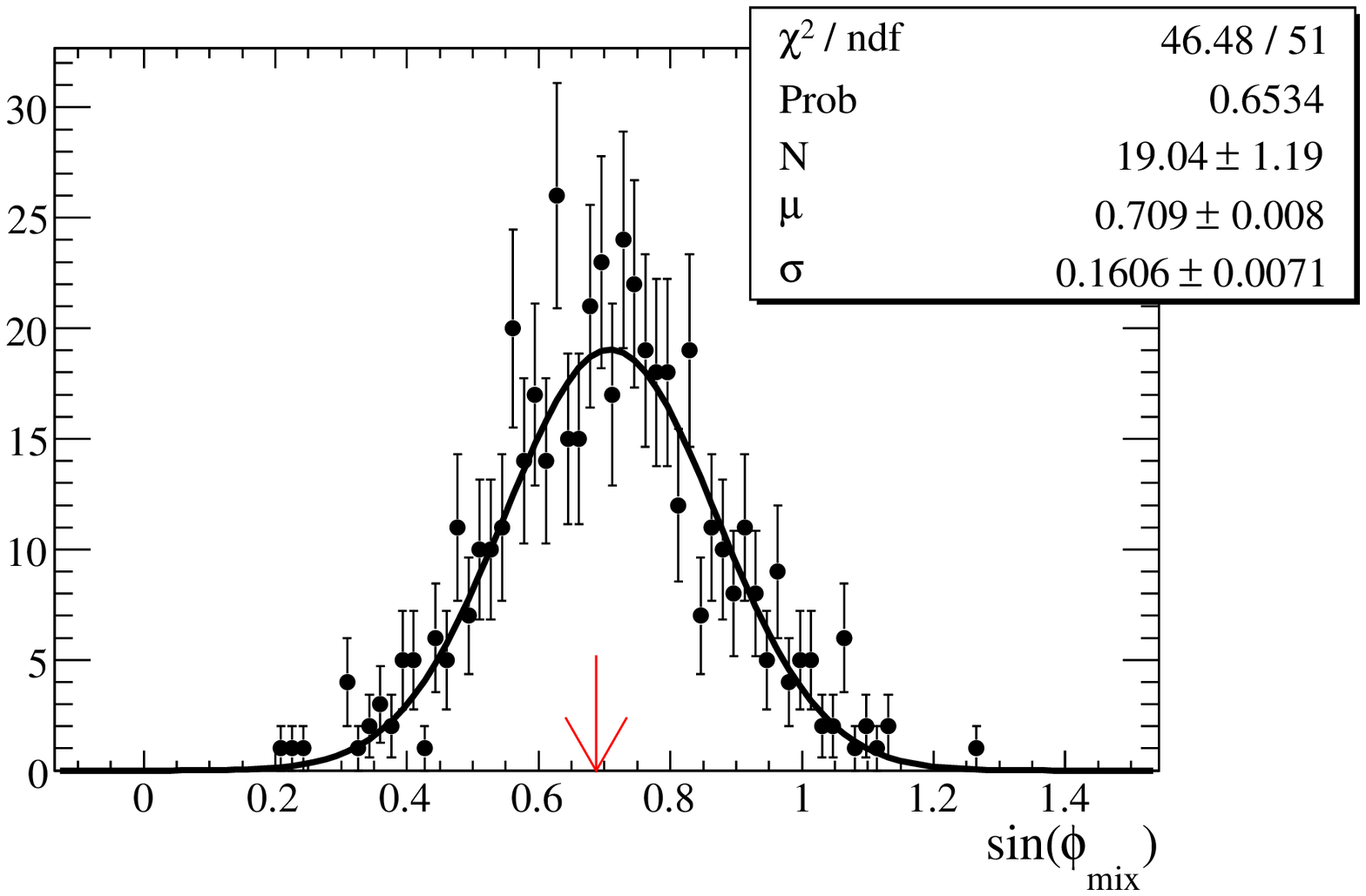}
  \caption{
    Distributions of the fitted values of (left) $\cos(\phi_{\rm mix})$ and
    (right) $\sin(\phi_{\rm mix})$ from the signal and background study.
    The resolutions are $0.54\pm0.02$ and $0.16\pm0.01$, respectively.
    The red arrows indicate the generated values, which are based on the
    value $2\beta = 21.7^\circ$.
  }
  \label{fig:fullFit}
\end{figure}

The results of this extended study are shown in Fig.~\ref{fig:fullFit}.
The fitted values of $\cos(\phi_{\rm mix})$ and $\sin(\phi_{\rm mix})$ are
still unbiased and the uncertainties are largely unchanged from the pure
signal case.

There are a number of additional potential experimental complications that we
have not simulated.  Effects due to smearing of the reconstructed Dalitz plot
position should be negligible since our Dalitz plot model does not contain any
very narrow resonances.  More significant misreconstruction effects can occur
when one of the particles from the signal side is exchanged with a particle
from the decay of the other $B$ in the event.  This effect tends to occur most
frequently near the corners of the Dalitz plot.  Since we veto the regions in
the corners of the Dalitz plot that are dominated by decays of the $D^*$
meson, we do not expect this to cause any serious difficulty.  Finally, we
have assumed that the reconstruction efficiency does not vary across the
$D \pi^+\pi^-$ phase space, but in a real experiment it is likely that the
corners of the Dalitz plot have a lower efficiency.  This may lead to a small
reduction in the precision of the $\cos(2\beta)$ measurement, since these are
the regions that are most sensitive to this parameter, as shown in
Fig.~\ref{fig:dp-amplitudes}.  However, in an experimental analysis the
efficiency would be measured using detailed Monte Carlo simulation together
with data control samples, and therefore we do not expect any bias on the
results.

\subsection{Prospects at LHCb}
\label{subsec:lhcb}

The potential to utilize this method at LHCb has not been explicitly studied,
but some useful extrapolations can be made.  
We consider only the $CP$-even $D$ meson decay to $K^+K^-$, which is
well-suited for study at LHCb since it has a final state consisting of only
charged tracks including two kaons.
(The use of particle identification information from LHCb's ring imaging
Cherenkov detectors being essential to reduce combinatoric background in the
hadronic environment~\cite{Alves:2008zz}.)

An estimation of the likely yield can be made by comparison with the decay
$B^0 \to D^- \pi^+$, $D^- \to K^+ \pi^-\pi^-$.  In this channel, which has a
product branching fraction of 
$(2.47 \pm 0.13) \times 10^{-4}$~\cite{Amsler:2008zz}, LHCb expects to trigger
and reconstruct $1.34 \times 10^6$ events in $2 \ {\rm fb}^{-1}$ of data (one
nominal year of data taking)~\cite{Gligorov:2007dp}.
For $B^0 \to D \pi^+\pi^-$, $D \to K^+K^-$ the product branching fraction is
$(3.3 \pm 0.4)\times 10^{-6}$ so that if the trigger and reconstruction
efficiencies are the same, one expects approximately 18,000 events.
Taking an effective tagging efficiency at LHCb of 5\%~\cite{Calvi:2007mw}, the
equivalent number of perfectly tagged events in one year of LHCb data is
approximately 900, which compares very well to the yields from the final BaBar
dataset (where the effective tagging efficiency is about
30\%~\cite{Aubert:2004zt}).
Although it may be necessary to apply less efficient selection criteria to
suppress background, this provides an indicative measure of the potential at
LHCb.  With $2 \ {\rm fb}^{-1}$ of data, it will be possible to achieve a
precision better than any previous measurements.  With the complete LHCb data
set, it should be possible to measure $\cos(2\beta)$ and $\sin(2\beta)$ with
precisions of $0.12$--$0.17$ and $0.03$--$0.05$ respectively.

The prospects for LHCb to make precise measurements of $\cos(2\beta)$ and
$\sin(2\beta)$ using $B^0 \to D \pi^+\pi^-$, $D \to K^+K^-$ therefore look
rather good.  However, any firm conclusion on this point requires a detailed
study including proper simulation of detector effects as well as consideration
of the Dalitz plot model including effects of the suppressed amplitudes.  We
leave such studies to further work.

\clearpage
\section{Conclusion}
\label{sec:conclusion}

We have presented the results of a feasibility study of a method to measure
$\cos(2\beta)$ using a time-dependent Dalitz plot analysis of 
$B^0 \to D \pi^+\pi^-$ decays, where the neutral $D$ meson is reconstructed in
decays to $CP$ eigenstates.
We estimate that, with the final BaBar dataset, $\cos(2\beta)$ can be measured
with a precision of $\sim 0.50$, making this approach competitive with, or
superior to, all other methods that have been attempted to date.
Furthermore, $\sin(2\beta)$ can be measured to within $\sim 0.16$,
which is more precise than any existing measurement using $b \to c \bar{u} d$
transitions.

We have argued that uncertainties relating to the composition of the 
$B^0 \to D \pi^+\pi^-$ Dalitz plot can be tamed using the much larger data
samples that are available when the $D$ is reconstructed in a flavour-specific
hadronic decay mode such as $D^0 \to K^- \pi^+$.  Therefore, the sensitivity is
limited by statistics only, and more precise measurements will be possible at
electron-positron colliders with higher luminosity.  Moreoever, since final
states containing only charged particles can be used, there is great potential
for this analysis to be employed at LHCb, where use of the $CP$-even decay 
$D \to K^+ K^-$ looks particularly promising.

Application of this method at future experiments will allow a determination of
$\cos(2\beta)$ that would definitively establish its sign, hence resolving the
ambiguity on $2\beta$. 
Moreover, measurements of $\sin(2\beta)$ in $b \to c \bar{u} d$ transitions
can be made with precision comparable to that obtained by the $B$ factories in
$b \to c \bar{c} s$ transitions ($B^0 \to J/\psi K^0$).
Discrepancies between these values would be unambiguous signs of new physics.

This work is inspired by the luminosity recorded by the $B$ factories,
and we are particularly grateful to our colleagues from PEP-II and BaBar.
We would like to thank Alex Bondar, Paul Harrison, Gagan Mohanty, 
Guy Wilkinson and Jure Zupan
for reading the manuscript and making useful suggestions.
This work is supported by the
Science and Technology Facilities Council (United Kingdom).

\bibliography{references}
\bibliographystyle{apsrev}

\end{document}